\documentclass[twocolumn,trackchanges]{aastex631}
\usepackage{CJKutf8}

\shorttitle{Quasar-associated \mbox{2175 \AA} Dust Absorbers}
\shortauthors{Zhang et al.}
 
\graphicspath{{./}}

\begin{document}
\begin{CJK}{UTF8}{gbsn}

\title{The Quasar-associated \mbox{2175 \AA} Dust Absorbers  
in the SDSS DR16 Quasar Catalog}

\author[0000-0001-8485-2814]{Shaohua Zhang}
\affiliation{Shanghai Key Lab for Astrophysics, Shanghai Normal University,  Shanghai 200234, China; zhangshaohua@shnu.edu.cn}

\author{Yuchong Luo}
\affiliation{Shanghai Key Lab for Astrophysics, Shanghai Normal University,  Shanghai 200234,  China; zhangshaohua@shnu.edu.cn}
\affiliation{National Astronomical Observatories, Chinese Academy of Sciences, 20A Datun Road, Beijing 100101, China}

\author{Shangchun Xie}
\affiliation{Shanghai Key Lab for Astrophysics, Shanghai Normal University,  Shanghai 200234, China; zhangshaohua@shnu.edu.cn}

\author{Chao Gao}
\affiliation{Shanghai Key Lab for Astrophysics, Shanghai Normal University,  Shanghai 200234, China; zhangshaohua@shnu.edu.cn}
\affiliation{Key Laboratory of Polar Science, Polar Research Institute of China, Ministry of Natural Resources, Shanghai 200136, China}

\author[0009-0009-1617-8747]{Zhijian Luo}
\affiliation{Shanghai Key Lab for Astrophysics, Shanghai Normal University,  Shanghai 200234, China; zhangshaohua@shnu.edu.cn}

\author{Chenggang Shu}
\affiliation{Shanghai Key Lab for Astrophysics, Shanghai Normal University, Shanghai 200234, China; zhangshaohua@shnu.edu.cn}

\author[0000-0001-8244-1229]{Hubing Xiao}
\affiliation{Shanghai Key Lab for Astrophysics, Shanghai Normal University,  Shanghai 200234, China; zhangshaohua@shnu.edu.cn}

\begin{abstract}
We present,  for the first time, a systematic study of quasar-associated 2175 \AA\ dust absorbers using spectroscopic data from the Sloan Digital Sky Survey  (SDSS)  Data Release 16 (DR16). 
By analyzing the optical spectra and multi-band magnitudes of 557,674 quasars in the redshift range of $0.7 \le z \le 2.4$, we identify 843  absorbers that share the same redshifts as quasars and are believed to originate from dust in the quasar nuclei, the host galaxies, or their surrounding environments.
These absorbers exhibit weak bump strengths ($A\rm_{bump}=0.49\pm0.15~\mu m^{-1}$) and narrow widths ($\gamma\rm=0.81\pm0.14~\mu m^{-1}$), while their peak positions span a broad range from $x_0 = 4.2$ to $4.84~ \mu m^{-1}$. 
Their average extinction curves resemble those of the Large Magellanic Cloud (LMC) but exhibit a shallower slope. 
In broad absorption line (BAL) quasars, the absorption bumps show systematic shifts in peak positions. Although further confirmation is needed, this may suggest environmental differences in dust grain properties.
We find a statistically significant negative correlation 
between bump strength and redshift, suggesting possible evolution in dust properties.
These findings highlight the changing composition and physical conditions of dust in quasar environments, likely influenced by factors such as metallicity, radiation fields, and dust processing mechanisms. 
Future studies incorporating ultraviolet and infrared data will be essential for refining the dust evolution models. 
Machine learning techniques and high-resolution spectroscopic follow-ups could enhance sample completeness and provide deeper insights into the chemical properties of the dust absorbers.

\end{abstract}


\section{INTRODUCTION}\label{sec:intro} 
A galaxy is a vast system composed of stars, stellar remnants, interstellar gas and dust, as well as dark matter, all bound together by gravity (\citealp{Sparke2007}).   
Although interstellar dust makes up only a small fraction of a galaxy's baryonic mass, it plays a critical role  in the structure and evolution of the galaxy.
Interstellar dust absorbs and scatters electromagnetic radiation emitted by celestial objects, altering their spectral energy distributions (SEDs). 
These processes, collectively referred to as interstellar extinction, significantly impact astronomical observations and the inferred physical properties of celestial objects.
Interstellar extinction varies with wavelength, with shorter wavelengths generally experiencing stronger extinction. 
The extinction magnitude at a given wavelength is represented by  $A_{\rm \lambda}$, and its variation with wavelength is referred to as the extinction law or extinction curve.
These extinction laws are important tools for understanding the chemical composition (e.g., \citealp{Li_Draine2001,Draine2003}) and size distribution (e.g., \citealp{Cardelli,Whittet2003}) of interstellar dust.
By studying the extinction laws of galaxies, astronomers can investigate the properties of interstellar dust and the surrounding medium of the galaxies, providing valuable insights into the chemical composition and star formation history of galaxies. 
However, our knowledge of dust extinction, particularly in high-redshift, remains limited.

The most notable and intriguing feature of the ultraviolet (UV) extinction curve is the broad absorption bump near \mbox{2175 \AA}, commonly referred to as the \mbox{2175 \AA} dust absorber.
This feature is widely observed in the Milky Way (MW) and Large Magellanic Clouds (LMC) \citep{Draine1989}, but is notably absent in starburst galaxies \citep{Calzetti1994} and the Small Magellanic Cloud (SMC) bar of active star formation \citep{Gordon1998}. 
It is believed  that strong star-formation activity destroys  the carriers of the \mbox{2175 \AA} dust absorber in these environments.
Studies of the MW extinction curve have shown  that the continuous linear components of extinction curves in different line-of-sight directions can be characterized by a single parameter $R_{\rm V}$ \mbox{($= A_V /E\rm(B - V )$)}, which is linked to the average size of dust particles in the interstellar medium \citep{Cardelli}.
However, the \mbox{2175 \AA} dust absorbers exhibit a constant central wavelength while their bandwidth varies: the peak position at \mbox{2175 \AA} remains remarkably stable, but the bandwidth differs from one line of sight to another \citep{Whittet2003}.
This variability suggests that the dust grains responsible for the \mbox{2175 \AA} dust absorber may differ from those contributing to the continuum linear component of the extinction curve.
The origin of the \mbox{2175 \AA} dust absorber, its environmental dependence, and its evolution have remained enigmatic for nearly six decades (e.g., \citealp{Stecher1965,Fitzpatrick1986,Draine2003}).
 Several potential carriers have been proposed to explain the absorption bump, including graphitic grains, amorphous silicates, organic carbon, and polycyclic aromatic hydrocarbon (PAH) molecules (e.g., \citealp{Draine2003,Xiang2011,Ma2020,Massa2022,Shivaei2022,Hirashita2023}). 

Directly measuring the extinction curve beyond the Local Group is challenging due to the  distance. 
Instead, more luminous background sources, such as quasars and gamma-ray burst (GRB) afterglows, are used to study interstellar gas and dust in distant galaxies.
Unlike observations in the Milky Way (MW), the \mbox{2175 \AA} dust absorber has been directly detected in only a few high-redshift sources.
A particularly rare and intriguing example is the gravitational lens system SBS 0909+532. 
In this case, the standard Galactic extinction law with $R_V=2.1\pm0.9$ was determined for the lensing galaxy (at redshift $z\rm_{lens}=0.83$) by analyzing the wavelength-dependent flux ratio between the images of the lensed quasar \citep{Motta2002}.
Most other detections of this feature are found in individual absorption systems along the lines of sight to background quasars. 
These include metal absorbers (e.g., \ion{Mg}{2}, \ion{Zn}{2}, \ion{C}{1}, and \ion{Ca}{2}; \citealp{Wang2004,Srianand2008,Noterdaeme2009,Jiang2010a,Jiang2010b,Jiang2011,Zhou2010,Zhou2022,Ledoux2015,Ma2017,Shi2020,Fang2023,Ge2024}) and Damped Lyman-$\alpha$ Absorbers (DLAs; e.g., \citealp{Wucknitz2003,Junkkarinen2004,Prochaska2009,Wang2012,Ma2015,Pan2017}). 
Similar detections have also been made toward six GRB afterglows, i.e.,
GRB 070802 \citep{Elasdttir2009}, GRB 080607 \citep{Prochaska2009}, GRBs 080605 and 080805 \citep{Zafar2012}, GRB 180325A \citep{Zafar2018} and GRB 140209A \citep{Greiner2024}.
These observations provide valuable insights into the properties of the interstellar medium (ISM) in the foreground galaxies of quasars, GRBs, or GRB host galaxies, rather than revealing information about quasars themselves.

In active galactic nuclei (AGNs), dust fills the interstellar space, forming a circumnuclear structure called the dusty torus--a critical element in AGN unified models \citep{Antonucci1993,Urry_Padovani1995,Netzer2015}.
Evidence for the dusty torus includes IR re-radiation of obscured light (e.g., \citealp{Nenkova2008,Alonso-Herrero2011}) and IR continuum bumps at a few microns (e.g., \citealp{Barvainis1987,Rodrguez-Ardila2006}).
Dust in AGNs is further confirmed by silicate absorption/emission bands at 10 and 18 microns (e.g., \citealp{Siebenmorgen2004,Hao2005,Spoon2007}) and polycyclic aromatic hydrocarbon (PAH) features at 3.3, 6.2, 7.7, 8.6, and 11.3 microns (e.g., \citealp{Siebenmorgen2004,Lutz2007,Farrah2010,Jensen2017,Esparza-Arredondo2018,Garca-Bernete2022,Garca-Bernete2004}). 
Despite this, studies of quasar composite spectra (e.g., \citealp{Richards2003,Czerny2004,Gaskell2004}), analyses of quasar color distributions (e.g., \citealp{Richards2003,Hopkins2004}), and individual source investigations (e.g., \citealp{Crenshaw2001,Gaskell2007}) 
indicate that quasar extinction curves typically resemble those of the SMC and lack evidence of the  \mbox{2175 \AA} dust absorber.
That is possibly due to the  grain destruction or growth in AGN environments (e.g., \citealp{Voit1992,Maiolino2001,Shao2017}).
Nevertheless, rare detections of \mbox{2175 \AA} dust absorbers associated with or intrinsic to quasars have been reported (e.g., \citealp{Gaskell2007,Jiang2011,Zhang2015,Zhang2022,Pan2017,Shi2020}).

In this work, we systematically searched for  quasar-associated \mbox{2175 \AA} dust absorbers at redshift \mbox{$z\sim 0.7 - 2.4$} in the spectra of the Sloan Digital Sky Survey IV (SDSS-IV) quasar program, i.e., the extended Baryon Oscillation Spectroscopic Survey\footnote{\url{https://www.sdss4.org/surveys/eboss/}} (eBOSS; \citealp{Dawson2016}).
The term ``quasar-associated" means that the \mbox{2175 \AA} dust absorbers share the same redshifts as quasars and originate from materials associated with the host or intrinsically in quasar.
The remainder of this paper is organized as follows: In Sect. 2 we describe the observational data available and data reduction. 
In Sect. 3 we present the derivation of the extinction curve, Sect. 4 provides a discussion, and Sect. 5 summarises the main conclusions.
Throughout this paper, we assume a $\Lambda$-dominated cosmology with $H\rm_0 = 72~ km~s^{-1}~Mpc^{-1}$, $\Omega\rm_M = 0.28$, and $\Omega\rm_\Lambda = 0.72$.

\section{Parent Sample}\label{sec:ps}
We based our analysis on the final quasar sample from the SDSS-IV eBOSS, i.e., the sixteenth released quasar catalog\footnote{\url{https://www.sdss4.org/dr16/algorithms/qso_catalog/}} (DR16Q; \citealp{Lyke2020}). 
This is currently the most comprehensive selection of quasars identified through spectroscopy to date, with an impressive completeness of 99.8\% with only a minimal contamination range of $0.3 - 1.3\%$.
The DR16Q catalog includes 750,414 quasars, including 99,856 broad absorption line (BAL) quasars and 35,686  DLA  quasars. 
All of these quasars have undergone a visual inspection to ensure reliable classifications and accurate redshift measurements.
The DR16Q catalog contains 557,674 quasars within the redshift range of interest ($0.7 \le z \le 2.4$), ensuring that the potential quasar-associated \mbox{2175 \AA}  bumps fall within the observed wavelength coverage of the BOSS spectrographs ($3560 - 10400$ \AA; Smee et al. 2013).
The catalog also integrates multiwavelength data from X-ray, ultraviolet (UV), optical, and near-/mid- infrared (NIR and MIR),  as well as  radio sky surveys. For more details, refer to Section 7 of \cite{Lyke2020}). 
We utilized  UV and infrared data in the Galaxy Evolution Explorer (GALEX; \citealp{Martin2005}), the Two Micron All Sky Survey (2MASS; \citealp{Skrutskie2006}), the UKIRT Infrared Deep Sky Survey (UKIDSS; \citealp{Lawrence2007}) and the Wide Field Infrared Survey Explorer (WISE; \citealp{Wright2010}) to expand the wavelength coverage of broad-band spectral energy distributions (SEDs) for more accurate extinction curves.
Thanks to its large sample size, high data quality, and well-distributed redshift range, the DR16Q dataset provides a valuable resource for a comprehensive exploration of and investigation into quasar-associated \mbox{2175 \AA} dust absorbers.

\section{Exploring Procedures} \label{sec:ep}
The pair approach is a widely used method for identifying  \mbox{2175 \AA} dust absorbers. 
This technique involves comparing observed spectra with those unaltered by dust absorption, and calculating the spectral ratio to establish the extinction curve \citep{Bless1970}.
In the MW, the comparison spectra are taken from two stars of the same spectral type, one of which has been reddened ($f_{\rm obs}$) and the other left unreddened ($f_{\rm model}$).
As a result, the extinction curve can be derived using the equation 
\begin{eqnarray} 
A(\lambda) = -2.5 \log \frac {f_{\rm obs}(\lambda)}  {f_{\rm model}(\lambda)}.
\end{eqnarray}
The Galactic extinction curves rise from the NIR to the near-ultraviolet (NUV), with a broad absorption feature at approximately $\lambda^{-1} \approx 4.6$ $\rm \mu m^{-1}$ ($\lambda \approx 2175$ \AA), followed by a sharp increase into the far-ultraviolet (FUV) with $\lambda^{-1} \approx 10$ $\rm \mu m^{-1}$.
The extinction curve in the optical and UV wavebands of the rest frame of the absorption system can be described by a parametric formula, which is of the form 
\begin{eqnarray} 
A(\lambda) = c_1 + c_2x + c_3 D(x, x_{0}, \gamma) + c_4 (x-c_5)^2, 
\end{eqnarray}
where $x~ (\equiv\lambda^{-1})$ is the inverse wavelength \citep{Fitzpatrick2007,Gordon2003}.
The first two terms represent the linear extinction component, determined by the slope $c_{2}$ and intercept $c_{1}$, which depend on the relative intensities of the model and observed spectra.
The third item incorporates a Drude profile, denoted as 
\begin{eqnarray}
D(x,x_{0}, \gamma) = \frac {x^2} {(x^2-x_0^2)^2+x^2\gamma^2}.
\end{eqnarray}
This profile is used to represent the 2175 \AA~bump. 
Here,  $x_{0}$ and $\gamma$ correspond respectively to the peak position and full width at half maximum (FWHM) of the Drude profile.
The strength of the bump can be described by the area under the Drude profile, $A_{\rm bump} = \int_{0}^{\infty} c_3 D(x,x_{0}, \gamma), \mathrm{d}x = \pi c_3 / 2\gamma$.
The fourth term represents the deviation from the extrapolated bump-plus-linear components, characterized by $c_4$ (the strength) and $c_5$ (the initial position of the second-order term) in the far-UV band at $x > c_5$.

The identification of  \mbox{2175 \AA} dust absorber in quasar spectra follows a method similar to that used for the MW. The observed quasar spectra are compared with either the quasar composite or an individual blue quasar spectrum (e.g., \citealp{Wang2004,Zhou2010,Jiang2011,Zhang2015,Zhang2022,Ma2017,Shi2020,Zhou2022}).
In this study, the quasar composite was created using optical spectra from 2200 SDSS quasar spectra \cite{VandenBerk2001} and 27 near-infrared (NIR) quasar spectra observed with the IRTF by \cite{Glikman2006}. 
The composite covers a wavelength range from 0.0860 to 3.52 $\mu m^{-1}$.
In Galactic extinction measurements, both reddened and unreddened stars maintain consistency in spectral type and luminosity.
However, this is not a case for quasar extinction measurements, where the observed and model spectra often differ.
Additionally, the SEDs and luminosities of quasars are intrinsically tied to the fundamental plane of black hole activity.
This necessitates an extinction model that accounts for variations in intrinsic spectral slopes and fluxes.
Therefore, $A(\lambda)$ represents an unnormalized relative extinction curve, and we cannot directly extract the conventionally defined extinction parameters (e.g., $A_{\rm V}$, $E(B-V)$, and $R_{\rm V}$). 
Specifically, if a quasar spectrum is bluer than the composite, the parameter $c_2$ may be negative, and $c_1$ will be zero when the composite is scaled to the observed spectrum in the infrared wavebands, where the extinction is negligible.
Nevertheless, since this study focuses on identifying  \mbox{2175 \AA} bumps, the bump features remain unaffected by intrinsic quasar variability.
Therefore, it is feasible to detect the bump from the derived extinction curve.
Similarly, the bump strength ($A_{\rm bump}$) is also unnormalized. 
In \cite{Gordon2003}  and \cite{Fitzpatrick2007}, the bump strength was defined as the normalized strength divided by the color excess ($A_{\rm bump}/E(B-V)$).

\begin{figure*}
\centering
\includegraphics[width=\hsize]{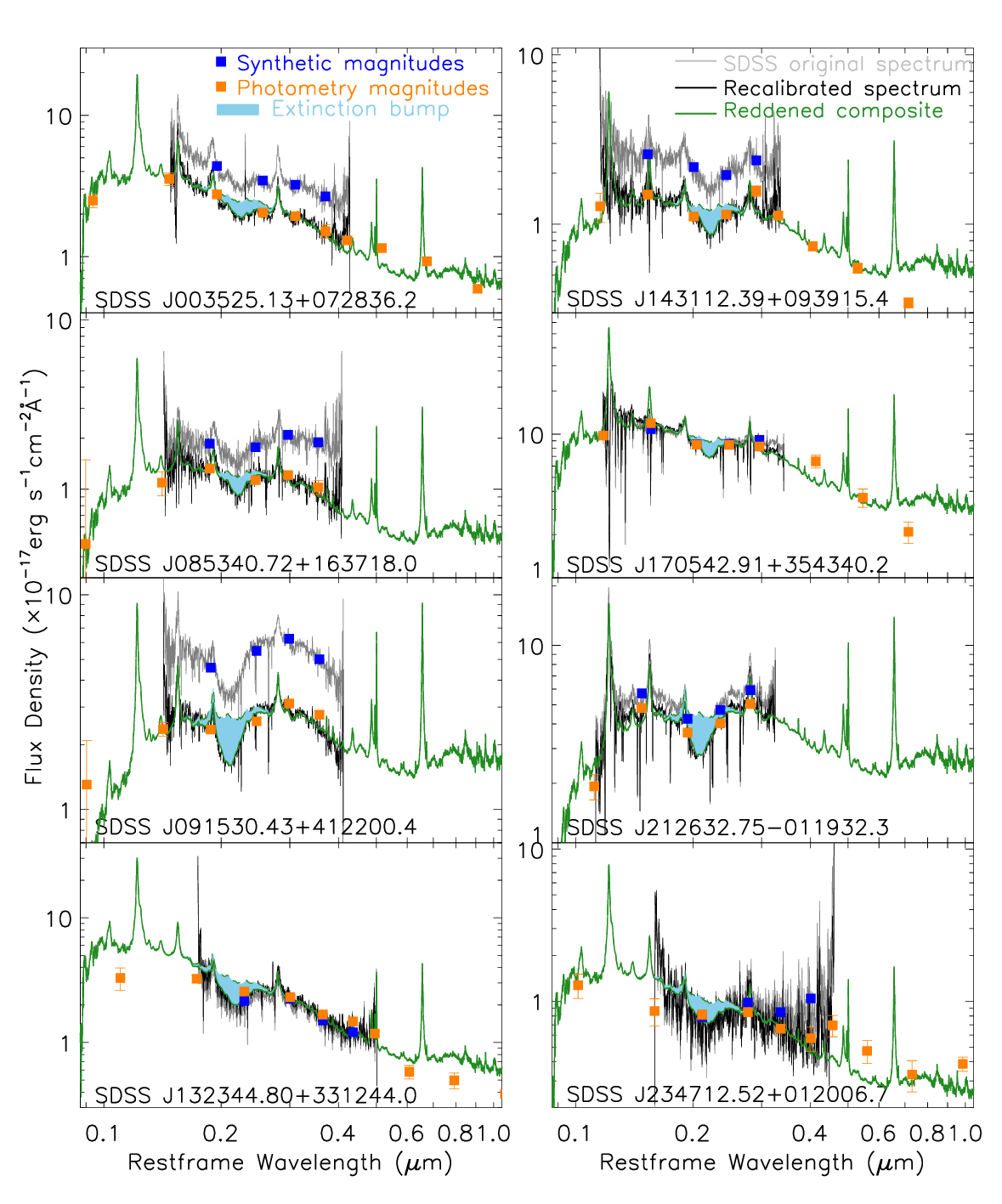}
\caption{Best-fit extinction models of eight quasars for examples. 
In each panel, the quasar spectra, including both the original SDSS spectrum (gray curve) and the recalibrated spectrum (black curve), are overlaid for comparison. 
Multi-band magnitudes from GALEX, SDSS, 2MASS, UKIDSS, and WISE are marked with orange squares, while  synthetic magnitudes at $g$, $r$, $i$, and $z$ bands, derived from the convolution of the original spectrum, are represented by blue squares. 
The solid and dashed green curves represent the reddened quasar composite, with or without the extinction bump component. 
The sky-blue region between these curves corresponds to the MW-like dust extinction bump.
The complete set of figures is available at \url{https://bit.ly/4g9lcO1}.}\label{fig1}
\end{figure*}

We derived the extinction curve for each quasar to identify quasar-associated \mbox{2175 \AA} dust absorbers.
First, we dereddened the observed spectroscopic and photometric data using the dust map from \cite{Schlegel1998}  and the reddening curve of  \cite{Fitzpatrick1999} to correct for Galactic extinction.
We then transformed the multiwavelength photometric data and spectroscopic fluxes into the quasar's rest frame using the redshifts suggested by \cite{Lyke2020}.
We then revisited SDSS spectrophotometric calibrations, addressing discrepancies between spectral fluxes and photometric data in some quasars.
In these cases, we computed the synthetic magnitudes at $g$, $r$, $i$, and $z$ bands by convolving the observed spectra, then fitted the flux ratios of the photometry and the synthetic magnitudes using a second-order polynomial. 
The resulting second-order correction curve was applied to each wavelength bin, adjusting the observed spectrum to match the photometry and obtaining a recalibrated spectrum for further analysis.
We then modeled the spectroscopic and photometric data using a composite reddened by the parameterized extinction curve. 
The fourth item was excluded from the parametric formula because the limited spectral wavelength coverage does not sufficiently constrain the extinction in the FUV band.
During fitting, we fixed the bump feature to the same redshift as quasar and allowed $x_0$ and $\gamma$ to vary within the Galactic and LMC curve parameters \citep{Gordon2003,Fitzpatrick2007}. 
We assigned higher weights to key regions, particularly the bump feature, to improve fitting accuracy.
Strong absorption lines and emission-like features were masked to avoid influencing the fitting of the  extinction and bump features.
To accurately model the extinction and bump curves, we adopted a three-stage fitting process, with each stage building upon the previous one in complexity, to optimize the fitting curve's parameters.
Least-squares minimization was performed using the Interactive Data Language (IDL) program MPFIT by C. Markwardt.
In the first two steps, we solved for the underlying extinction and bump components: fixing the weights of the bump region to zero and solving for $c_1$ and $c_2$, then fixing those values and solving for $c_3$, $x_0$, and $\gamma$. 
This approach ensured a better fit with only two or three free parameters (extinction component).
In the third step, using these values as initial guesses, we rejected offsets greater than $3\sigma$ between the observed data and the reddened composite and refitted the data. 
We repeated this process until the $\chi^2$ value converged and the offsets were rejected according to the latest fitting iteration.
In Figure \ref{fig1}, we present the observed data and the best-fit models for eight  quasars.

From our initial search, we identified preliminary candidates for the bump feature with a positive $c_3$ value in nearly 40\% of quasars.
\cite{Pitman2000} pointed out that the variability of broad \ion{Fe}{2} emission multiplets can resemble an extinction bump in a quasar spectrum, especially in the context of quasar-associated \mbox{2175 \AA} dust absorbers.
These absorbers are often located to the left of the quasar's small blue bump. 
However, many of the preliminary bump candidates might be false positives, primarily due to broad iron emission lines.

To assess the significance of the candidates and exclude false positives, we applied a simulation technique based on control samples. Developed by \cite{Jiang2010a,Jiang2010b}, this technique has been successfully used in previous searches for bump features in quasar spectra (e.g., \citealp{Jiang2011,Ma2015,Zhang2015,Zhang2022}).
The simulation begins by constructing a quasar spectral control sample with redshifts similar to those of the candidate. 
We then fit each spectrum in the control sample using a reddening curve parameterized, setting the parameters $x_0$ and $\gamma$ based on the optimal values derived from the candidate's bump feature.
We collected the best-fitting bump strengths from the control sample, and the distribution of these strengths reflects random fluctuations in the quasar continuum and variations in broad \ion{Fe}{2} emission multiplets, assuming no natural extinction bump is present. 
This strength distribution is well-described by a Gaussian profile. If the candidate bump deviates significantly from this distribution, it is considered statistically significant.
In this work, for each bump candidate, we selected control quasar spectra with an $i'-$band signal-to-noise ratio of $S/N \ge 6$ and emission redshifts within the range $z_{\rm em} - 0.05 < z < z_{\rm em} + 0.05$, where $z_{\rm em}$ is the emission redshift of the quasar from the DR16 database.
Only bumps with significance greater than 3$\sigma$ were considered as effective detections. 

\begin{deluxetable*}{cccrcccc}
\tablecaption{Quasar-associated \mbox{2175 \AA} Dust Absorbers in SDSS DR16 Quasar Catalog\label{table1}}
\tablehead{
\colhead{Name} & \colhead{z} & \colhead{$c_1$} & \colhead{$c_2$} & \colhead{$c_3$} & \colhead{$x_0$} & \colhead{$\gamma$} & \colhead{Significane}  \\ 	\colhead{SDSS J}    & \colhead{ } & \colhead{(mag)} & \colhead{(mag)} & \colhead{(mag)} & \colhead{($\mu$m$^{-1}$)} & \colhead{($\mu$m$^{-1}$)} & \colhead{ }} 
\startdata
000402.91+221434.1 & 1.3140 & 8.07$\pm$0.03 & 0.23$\pm$0.01 & 0.19$\pm$0.04 & 4.44$\pm$0.04 & 0.81$\pm$0.12 &3.80$\sigma$  \\
000526.01+110109.1 & 1.9219 & 8.28$\pm$0.02 & 0.20$\pm$0.01 & 0.21$\pm$0.03 & 4.61$\pm$0.02 & 0.82$\pm$0.06 &5.11$\sigma$ \\
000649.21+244047.5 & 1.8598 & 9.47$\pm$0.03 & 0.01$\pm$0.01 & 0.25$\pm$0.04 & 4.25$\pm$0.02 & 0.84$\pm$0.06 &7.80$\sigma$ \\
000756.54+192432.6 & 1.6450 & 8.61$\pm$0.01 &-0.04$\pm$0.01 & 0.15$\pm$0.01 & 4.20$\pm$0.01 & 0.69$\pm$0.03 &4.87$\sigma$ \\
000929.39+240126.6 & 2.3860 & 9.40$\pm$0.04 & 0.03$\pm$0.01 & 0.14$\pm$0.02 & 4.53$\pm$0.02 & 0.54$\pm$0.05 &4.75$\sigma$ \\
000956.82+224450.1 & 1.9409 & 7.71$\pm$0.01 & 0.06$\pm$0.01 & 0.28$\pm$0.02 & 4.43$\pm$0.02 & 0.99$\pm$0.06 &3.29$\sigma$ \\
001132.99+241356.5 & 1.2342 & 7.95$\pm$0.01 &-0.17$\pm$0.01 & 0.29$\pm$0.01 & 4.48$\pm$0.01 & 0.96$\pm$0.03 &3.45$\sigma$ \\
001140.23+165905.1 & 2.1150 & 8.66$\pm$0.03 & 0.11$\pm$0.01 & 0.18$\pm$0.02 & 4.45$\pm$0.02 & 0.73$\pm$0.06 &4.59$\sigma$ \\
001413.76+121146.9 & 1.2010 & 7.19$\pm$0.03 & 0.37$\pm$0.01 & 0.38$\pm$0.05 & 4.70$\pm$0.03 & 0.95$\pm$0.08 &5.57$\sigma$ \\
001433.17+194516.2 & 1.5900 & 8.30$\pm$0.02 & 0.20$\pm$0.01 & 0.40$\pm$0.05 & 4.77$\pm$0.04 & 0.97$\pm$0.09 &3.08$\sigma$ \\
\enddata
\tablecomments{ An complete version of this table is available online.}
\end{deluxetable*}

\begin{figure}
\centering
\includegraphics[width=\hsize]{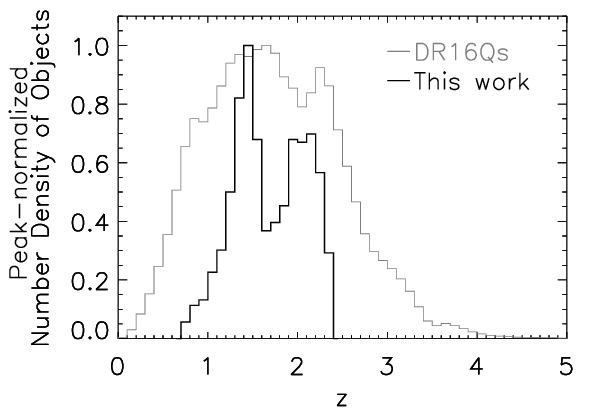}
\caption{Redshift distribution of quasar-associated \mbox{2175 \AA} dust absorbers (black curve), with that of DR16Q quasars (grey curve) overplotted for comparison.} \label{fig2}
\end{figure}

After visually checking the data and removing false signals or heavily trimmed spectra, we identified 843 quasar-associated \mbox{2175 \AA} dust absorbers (see Table \ref{table1}).
Figure \ref{fig2} shows the redshift distribution of these dust absorbers, with the redshift distribution of DR16Q quasars included for comparison.
Notably, the detection rate of quasar-associated \mbox{2175 \AA} dust absorbers is lower at $z<1.2$. This may be due to the limited wavelength coverage of the BOSS spectrographs and the lower detection efficiency in UV photometry. Expanding the GALEX survey to include more NUV and FUV data for quasars could help identify more \mbox{2175 \AA} dust absorbers at lower redshifts.
Another unexpected feature appears around $z \sim 1.7$, where the numbers of dust absorbers drops to just over half of that at higher or lower redshifts. 
It remains unclear whether this is a real effect or a bias introduced by the spectral fitting and bump identification process. 
Theoretically, the rest-frame spectrum at this redshift is well covered by the BOSS spectrographs, with the available wavelength range effectively capturing both the bump absorption region and the adjacent unabsorbed regions on either side.

In \cite{Zhang2015}, we quantitatively investigated the potential impact of broad iron emission/absorption on detecting the \mbox{2175 \AA} dust absorbers. 
We constructed a simulated quasar sample with varying \ion{Fe}{2} emission strengths and used this sample to search for the bump feature at \mbox{2175 \AA} with different widths.
The influence of \ion{Fe}{2} emission on bump detection increases with emission strength but diminishes as the absorber width increases.
We also derived the detection threshold for this simulated sample using the same procedure described earlier and found that almost all false ``bumps"  in the simulated spectra were below the $3\sigma$ threshold. 
The only exceptions were quasars with exceptionally strong iron emission, such as those with iron emission strengths three times higher than the average in the SDSS quasar catalog. For more details, please refer to Section 3 of \cite{Zhang2015}.
In Section 4 of \cite{Zhang2015}, we also ruled out the possibility that the \mbox{2175 \AA} dust absorbers are simply an illusion caused by unusual \ion{Fe}{2} and/or \ion{Fe}{3} absorption troughs, using photoionization simulations of iron absorption. 
There are not enough iron transitions near \mbox{2175 \AA} to create wide and strong enough absorption troughs to mimic the \mbox{2175 \AA} dust absorbers.

\begin{figure*}
	\centering
	\includegraphics[width=\hsize]{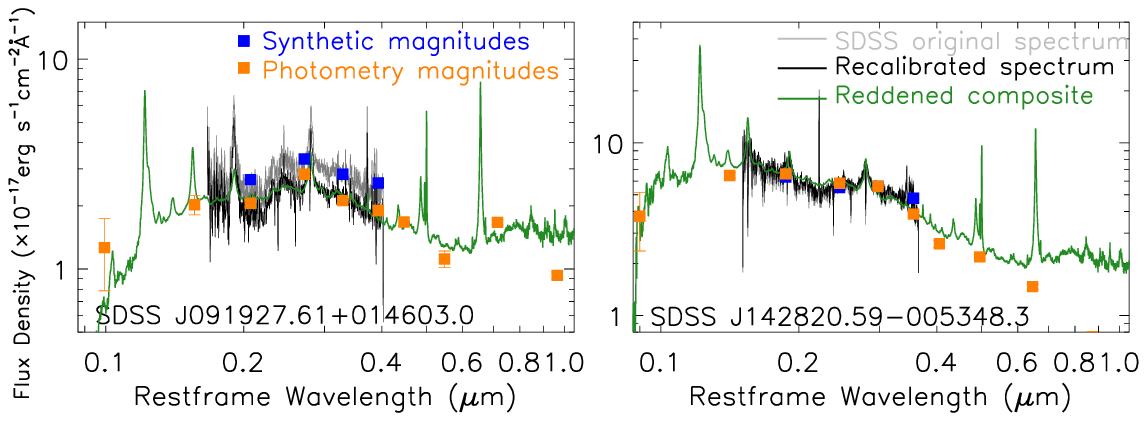}
	\caption{Best-fit extinction models for two rejected candidates reported by Jiang et al. (2011). The annotations are the same as those in Figure \ref{fig1}. } \label{fig3}
\end{figure*}

In previous studies, a total of 18 \mbox{2175 \AA} dust absorbers, either associated with or intrinsic to quasars, have been identified, 12 of which are included in our sample. 
\cite{Gaskell2007} reported that Mrk 304 shows an LMC-like extinction curve, likely including a \mbox{2175 \AA} bump. 
However, its BOSS spectrum does not cover the bump region due to its low redshift ($z = 0.0663$).
\cite{Jiang2011}  identified 39 \mbox{2175 \AA} dust absorber candidates associated with strong \ion{Mg}{2} absorption lines in quasar spectra from SDSS DR3 at $z \sim 1-1.8$. 
Among these, five candidates exhibit a velocity offset between the \ion{Mg}{2} absorption lines and the quasar rest frame of $v_{\rm off} \leq 5,000~ \rm km/s$, suggesting that these \mbox{2175 \AA} dust absorbers are quasar-associated.
However, two candidates are rejected in this work, 
as shown in Figure \ref{fig3}. Although weak \mbox{2175 \AA} bump features are observed, they do not meet the $3\sigma$ significance threshold.
The \mbox{2175 \AA} dust absorbers reported by \cite{Zhang2015}, \cite{Pan2017} and \cite{Shi2020} are all reconfirmed in this work. 
In contrast, three \mbox{2175 \AA} dust absorbers in FeLoBAL quasars from \cite{Zhang2022} are missed.
Due to the presence of overlapping iron absorption, it is necessary to manually identify and mask the wavelength ranges of \ion{Fe}{2} troughs during spectral fitting and bump detection. The fitting method employed in this work is not applicable to such cases.

\section{Results and Discussion}
\begin{figure*}
\centering
\includegraphics[width=\hsize]{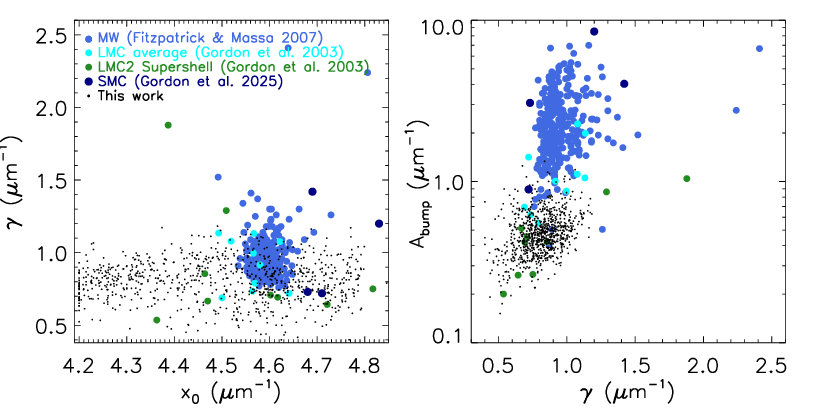}
\includegraphics[width=\hsize]{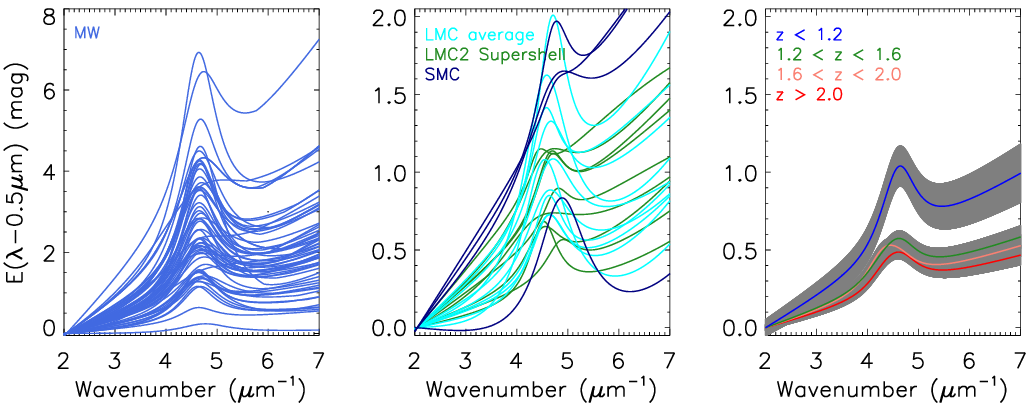}
\caption{Top: Comparison of bump parameters with \mbox{2175 \AA} bumps in MW (blue circles), LMC average (cyan circles), LMC2 Supershell (green circles) and SMC (navy circles).
The quasar-associated \mbox{2175 \AA} dust absorbers are labeled with black points.
Bottom: Colored plots display the average extinction curves with errors of quasar-associated \mbox{2175 \AA} dust absorbers at four different redshift bins. The extinction curves in MW, LMC average, LMC2 Supershell and SMC are also plotted for comparison.} \label{fig4}
\end{figure*}

Top panels of Figure \ref{fig4} present the bump parameters of quasar-associated \mbox{2175 \AA} dust absorbers. 
For comparison, we include measurements from the Local Group: MW data in blue \citep{Fitzpatrick2007}, LMC average in cyan and LMC2 Supershell in green \citep{Gordon2003}, and SMC in navy \citep{Gordon2024}.  
MW data show that the \mbox{2175 \AA} dust absorbers exhibit strong absorption features, with a bump strength of $A_{\rm bump} = 2.48 \pm 1.15~\rm \mu m^{-1}$, a peak position of $x_0 = 4.59 \pm 0.03~\rm \mu m^{-1}$, and a significantly variable width of $\gamma = 0.95 \pm 0.158~\rm \mu m^{-1}$.
In contrast, the LMC average and LMC2 Supershell dust absorbers exhibit a much wider range of parameter distributions.
The peak positions of \mbox{2175 \AA} bumps range from $x_0 = 4.36$ to $4.81~\rm \mu m^{-1}$, widths vary from $\gamma = 0.54$ to $1.88~\rm \mu m^{-1}$, and strengths span $A_{\rm bump} = 0.2$ to $2.27~\rm \mu m^{-1}$.
Notably, the \mbox{2175 \AA} bump is significantly weaker in the LMC2 Supershell, whereas the rest of the LMC average follows a curve similar to the average Galactic extinction curve.
Additionally, the SMC extinction curves fall into two categories: the traditional curve, which lacks a \mbox{2175 \AA} bump, and the others feature a recognizable \mbox{2175 \AA} bump. 
\cite{Gordon2024} identified 4 sightlines among 32 in the SMC that show MW-like extinction with a detectable \mbox{2175 \AA} bump, they tend to have larger $x_0$ values and stronger bump strengths even than those in the MW.

The quasar-associated \mbox{2175 \AA} dust absorbers exhibit significant differences compared to those in the Local Group. 
In particular, while their peak positions generally fall within the same parameter space as the LMC average, LMC2 Supershell, and SMC absorbers, a substantial fraction clusters within $x_0 = 4.2$ to $4.36~\rm \mu m^{-1}$.
However, as $x_0 = 4.2~\rm \mu m^{-1}$  is the lower limit in the fitting process, this clustering may not fully reflect the true distribution.
This is likely due to the presence of BAL quasars, which constitute more than one-third of the sample. For these sources, the bumps tend to be present toward lower $x_0$ values. 
For the other parameters, the mean bump strength ($A_{\rm bump} = 0.49 \pm 0.15~\rm \mu m^{-1}$) and width ($\gamma = 0.81 \pm 0.14~\rm \mu m^{-1}$) are approximately one-fifth or slightly smaller than the corresponding values observed in the MW, respectively.

Although the exact nature of the \mbox{2175 \AA} dust absorbers remains uncertain, the width and position of the bump offer valuable clues. 
PAH molecules, which exhibit strong $\pi \to \pi\ast$ electronic transitions with primary absorption bands in the $200 - 250$ nm range, it is therefore natural to attribute at least part of the 2175 \AA\ extinction feature to PAHs \citep{Li_Draine2001}.
\cite{Gordon2024} found the 2175 \AA\ bump area is correlated with the mass fraction of PAH grains $q_{\rm PAH}$ in the combined SMC and LMC samples.
\cite{Draine2003} suggested that variations in the width and slight shifts in the peak position of the Galactic 2175 \AA\ extinction bump could be attributed to differences in the PAH mixture. 
Similarly, the slightly narrower width and the widely varying of peak positions observed in quasar-associated \mbox{2175 \AA} dust absorbers suggest that the PAH composition in quasar environments differs significantly from that in local galaxies. 
\cite{Lin2023} performed quantum chemical computations on the electronic transitions of 30 compact PAH molecules, finding that the peak position shifts to longer wavelengths as PAH size increases.
This also raises a question: Could the size of PAH molecules in BALQs be larger and contain more carbon atoms?
On the other hand, the PAH emission features between 3 and 12 microns have been well studied and detected in both galaxies and quasars (e.g., \citealp{BregmanTemi2005,Farrah2010}).
Indeed, a correlation between the \mbox{2175 \AA} hump and the IRAS 12 $\mu$m emission (dominated by PAHs) was found by \cite{Boulanger1994} in the Chamaeleon molecular complex, and \cite{Massa2022} reported that three PAH emission features (7.6, 8.5, and 11.3 $\mu$m) are correlated with the 2175 \AA\ bump.
Follow-up MIR spectroscopic observations will be crucial for further constraining these differences. By distinguishing between various dust emission features, such observations can provide deeper insights into the composition and evolution of dust in quasars.

In the bottom panels of Figure \ref{fig4}, we present a comparison between the measured average extinction curves of quasar-associated \mbox{2175 \AA} dust absorbers and those of the Local Group. 
As discussed in the previous section, although determining the absolute extinction curve for an individual quasar remains observationally challenging, 
we can statistically derive the average extinction curve of our absorbers (the derived extinction curves are setting the extinction to zero for $\lambda \to \infty$) by subtracting that of their corresponding control sample quasars. 
We categorize the \mbox{2175 \AA} dust absorbers into four redshift bins at  \(z\rm_{abs} = 1.2,\ 1.6, \  and \  2.0\),  and the resulting average extinction curves for each bin are plotted with distinct colors, along with their associated uncertainties.
The overall slopes are similar to the LMC average and LMC2 curves, but they are somewhat shallower, which  appears to be consistent with the independent measurements from UV selected galaxies and star-forming galaxies at similar redshifts (\citealp{Noll2007,Buat2012,Kriek_Conroy2013,Scoville2015}).

In the DR16Q catalog, the BAL probability is provided as a fundamental parameter for each quasar, derived from the statistical significance of the absorption troughs associated with \ion{C}{4}.
Through  automated algorithms, nearly 100,000 BAL quasars have been identified and cataloged.
Similarly, \cite{Anand2022} employed an automated absorption detection pipeline to identify approximately 160,000 narrow \ion{Mg}{2} absorbers\footnote{\url{https://wwwmpa.mpa-garching.mpg.de/SDSS/MgII/}} within  the DR16Q sample.
A thorough  visual inspection was conducted on the spectra of 843 sources in our sample. 
We classified 145 sources as LoBALQs, characterized by the presence of \ion{Al}{3} and/or \ion{Mg}{2} BALs, and 155 sources as HiBALQs, exhibiting only \ion{C}{4} BALs.
In other words, more than one-third of the sources are BAL quasars--a result consistent with our previous work \citep{Zhang2015,Zhang2022}--and significantly higher than the fractions observed in optically selected quasar samples (e.g., \citealp{Weymann1991,Gibson2009,Zhang2010}).
For NAL systems, we classified  160 as quasar-associated, as their velocities fall within 5000 km/s of the quasar redshift. 
Additionally, we identified 95 systems as high-velocity NALs (commonly termed  intervening systems and labeled here as ``HvNALs"), characterized by significantly larger blueshifted velocities, typically originating from foreground galaxies of quasars.
Among the quasar-associated systems, absorbers are detected on the red side of the \ion{Mg}{2} and/or \ion{C}{4} emission line centers in 35 sources (labeled as ``RNALQs") and on the blue side in 124 sources (labeled as ``BNALQs").
For the remaining 289 sources, no significant broad or narrow absorption systems are identified in their spectra, based on the current spectral $S/N$.

\begin{figure*}
	\centering
	\includegraphics[width=\hsize]{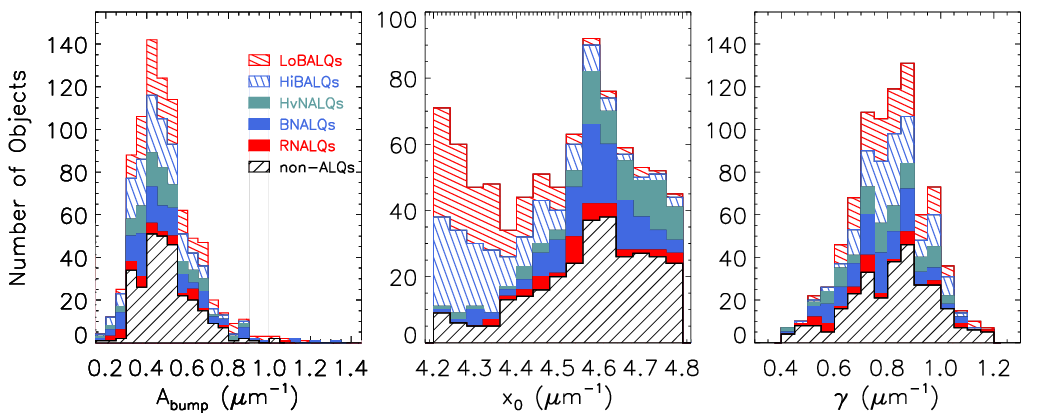}
	\includegraphics[width=\hsize]{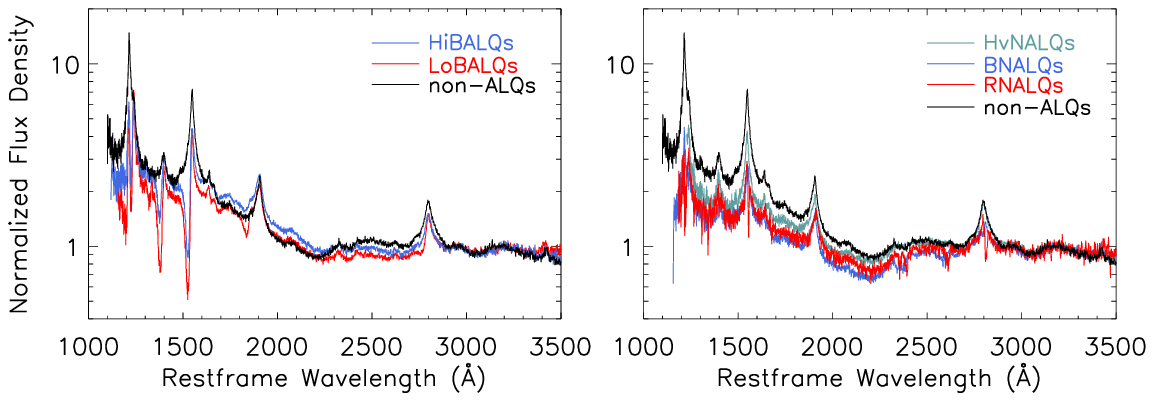}
	\caption{Comparison of bump parameter distributions (top panels) and composite spectra (bottom panels) of quasars with and without BALs and NALs.} \label{fig5}
\end{figure*}   

Top panels of Figure \ref{fig5} display  the distribution of bump parameters, categorized accordingly. 
No significant differences are observed in the bump strength or width  across the various types.
Regarding the peak position, both HiBALQs and LoBALQs show a tendency toward  smaller $x_0$ values. 
This seems to suggest that the presence of \ion{C}{4} and \ion{Mg}{2} BAL troughs cause the 2175 \AA\ bumps to shift closer to \ion{Mg}{2} emission line.
These findings are further corroborated by the geometrically composite spectra displayed in the bottom-left panel of Figure \ref{fig5}.
For clarity, the composite spectra have been scaled to normalized  at 3200 \AA.
If absorption troughs were indeed responsible for this effect, a significant increase in the bump width would be expected.
However, in the case of HiBALQs, no absorption troughs are detected between the \ion{Al}{3} and \ion{Mg}{2} emission lines.
Additionally, spectral fitting results confirm that BAL troughs do not affect the bump fitting.
This suggests  that the \mbox{2175 \AA} dust absorbers in BALQs may differ from those in non-BAL quasars, particularly in peak positions.
A possible explanation is that the average dust grain size responsible for \mbox{2175 \AA} dust absorbers differs between these two environments.

In HvNALQs, the \mbox{2175 \AA} dust absorbers typically exhibit larger average $x_0$ values. 
If  \mbox{2175 \AA} dust absorbers are associated with these high-velocity NAL systems while their redshifts are tied to those of  quasars during analysis, a larger $x_0$ would naturally be expected.
It is worth noting that not all \mbox{2175 \AA} dust absorbers in HvNALQs originate from the foreground galaxies. 
In fact, these are simply intervening absorption line systems present in the quasar spectra, and it remains unclear whether these \mbox{2175 \AA} dust absorbers are physically linked to NAL systems.
For both redshifted and blueshifted NALQs, the peak positions and widths of \mbox{2175 \AA} dust absorbers are consistent, with stronger bumps observed in BNALQs.
As shown in the bottom-right panel, the composite spectra of RNALQs and BNALQs are nearly identical, except in the bump region. Additionally, all composite spectra of NAL quasars are, on average, redder than that of non-ALQs, indicating stronger extinction of the continuum.

In previous studies, nearly all \mbox{2175 \AA} dust absorbers have been observed to be associated with metal absorption lines and (sub-)DLA systems.
This naturally raises the question of how \mbox{2175 \AA} dust absorbers related to these absorption systems and other physical properties. 
Follow-up high-resolution spectroscopy has shown that some \mbox{2175 \AA} dust absorbers have high metallicity (half of them with super-solar metallicity) and heavy dust depletion (e.g., \citealp{Noterdaeme2009,Wang2012,Ma2017,Ma2018,Pan2017}). 
The majority of these absorbers display dust depletion levels comparable to those found in the Milky Way's diffuse clouds.
Furthermore, a correlation between bump strength and dust depletion appears to exist.
In our sample, 30 percent of the sources exhibit strong absorption lines, making it essential to conduct follow-up observations.
A detailed analysis of these absorption systems will provide independent measurements of chemical abundances, dust depletion levels, total hydrogen column densities, and dust-to-gas ratios. These measurements will be crucial for identifying the optimal conditions under which 2175 \AA\ dust absorber carriers exist.

\begin{figure*}
\centering
\includegraphics[width=0.9\hsize]{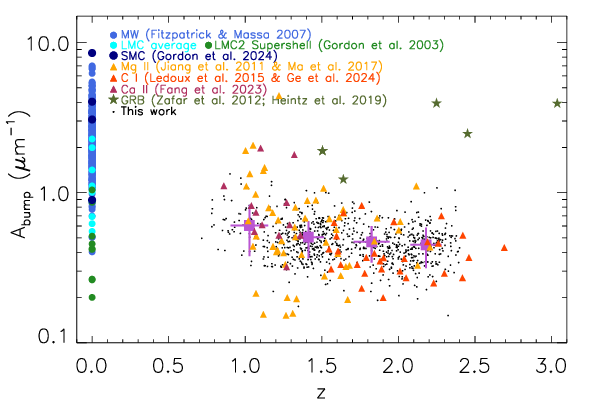}
\caption{Redshift vs. \mbox{2175 \AA} bump strength for different dust absorber populations. 
Quasar-associated \mbox{2175 \AA} dust absorbers are represented by black points. 
Blue circles indicate dust bumps observed in the MW, navy circles denote bumps in the SMC, while cyan and green circles correspond to those in the LMC average and LMC2 supershell, respectively.  
Orchid squares represent the average values at different redshifts, revealing a general trend of decreasing \mbox{2175 \AA} bump strength with increasing redshift. 
For comparison, \mbox{2175 \AA} dust absorbers associated with other absorption systems and GRB afterglow are also shown:  
Red triangles represent dust bumps detected in \ion{C}{1} absorbers, orange triangles correspond to those found in \ion{Mg}{2} absorbers, while maroon triangles mark bumps identified in \ion{Ca}{2} absorbers.  
Olive stars denote \mbox{2175 \AA} bumps observed in GRB optical afterglow.} \label{fig6}
\end{figure*}

Figure \ref{fig6} shows the distribution of bump strength as a function of redshift for quasar-associated \mbox{2175 \AA} dust absorbers.
The most notable feature of this plot is the observed decline in bump strength with increasing redshift, despite significant local scatter—similar to that seen in the MW. 
This evolutionary trend is statistically significant, with a Spearman's correlation  yielding a null hypothesis probability of \( P_r \sim 10^{-10} \), indicating that  grains may have grown in \mbox{2175 \AA} dust absorbers at $z\sim 0.7 - 2.4 $. 
This also appears to be consistent with the findings for UV-selected galaxies (\citealp{Noll2007,Buat2012}), but at much higher sensitivity due  to the relative high flux of quasars.
The orchid-colored squares represent the mean values in four redshift bins, which follow a trend line with a slope of \(-0.13\).
 
In our ongoing investigation, we have conducted a systematic search for 2175 \AA\ dust absorbers associated with metal absorption lines in the spectra of SDSS DR16 quasars. Our initial identified over 2,000 candidates, the majority of which are intervening absorbers (Zhang et al., in prep.).
When compared to the quasar-associated sample in this study, these absorber candidates cover a redshift range from 0.7 to 2.4. Notably, nearly one-third of the sources have redshifts below 1.0, while less than one-fifth exceed a redshift of 1.5. Additionally, these absorber candidates exhibit significantly stronger bump strengths, with an average value of approximately $\rm 0.85 \pm 0.51 ~\mu m^{-1}$. 
This  sample partially bridges the gap between the quasar-associated sample and the MW bumps, both in terms of redshift distribution and bump strength.
When combining both datasets, the evidence indicates a clear trend: bump strength intensifies as redshift decreases, eventually reaching the value measured in the MW. This trend appears to track the chemical enrichment of the ISM. 

Additionally, we include \mbox{2175 \AA} dust absorbers detected in strong quasar absorption systems (\citealp{Jiang2011, Ledoux2015,Ma2017,Fang2023,Ge2024}) and GRB afterglows (\citealp{Zafar2012,Heintz2019}). 
Excluding the five quasar-associated absorbers from \cite{Jiang2011}, all other absorbers originate from the galactic environments, exhibiting a strength distribution generally consistent with our results. 
However, a few outliers deviate from this trend.  
Several \mbox{2175 \AA} dust absorbers with very weak bump strengths are classified as low-confidence candidates in the Jiang et al. sample. 
Conversely, nearly twenty sources exhibit exceptionally strong bump strengths, comparable to the MW \mbox{2175 \AA} bumps. 
However, such strong bumps are absent in quasar-associated \mbox{2175 \AA} dust absorbers.
This does not necessarily mean that strong bumps are entirely absent in these absorbers.
Dust extinction and reddening can change the magnitude and color of quasars, potentially affecting their detection. 
In particular, heavily reddened quasars may be missed by the SDSS quasar target selection algorithm \citep{Richards2002}.
With rapid advances in computing, machine learning has become a powerful tool for classification and feature detection in astronomy.
To overcome the limitations of using quasar spectral data to search for 2175 \AA\ dust absorbers, a more effective method is to use computational techniques to directly identify these absorbers in SDSS multi-band magnitude data. Combining this with efficient follow-up low-resolution spectroscopy will greatly improve sample completeness.

\section{Summary}
In this paper, we present, for the first time, a systematic study of  quasar-associated \mbox{2175 \AA} dust absorbers using spectroscopic data from the SDSS-IV DR16. These absorbers, which share the same redshift as their background quasars, are believed to originate from dust within the quasar nuclei, the host galaxies, or even their surrounding environments.
We identify 843 \mbox{2175 \AA} dust absorbers at a high confidence level in the redshift range of \(0.7 \leq z \leq 2.4\), expanding the existing sample size by nearly 50 times.

Quasar-associated \mbox{2175 \AA} dust absorbers exhibit weaker bump strengths ($A\rm_{bump}=0.49\pm0.15~\mu m^{-1}$) and narrower widths ($\gamma\rm=0.81\pm0.14~\mu m^{-1}$), while their peak positions span a broad range from $x_0 = 4.2$ to $4.84~ \mu m^{-1}$. 
The relative extinction curves derived from spectral fitting and their average extinction curves suggest that these absorbers share similarities with LMC dust, but with a slightly shallower slope. 
Within this sample, over a third of the sources are BAL quasars, a fraction that significantly exceeds that found in normal optically selected quasars. This suggests a potential connection between the \mbox{2175 \AA} dust absorbers and the BAL systems.

The most striking finding is the decline in bump strength with increasing redshift.
This trend is statistically significant, as indicated by a Spearman's correlation, which yields a null hypothesis probability of \( P_r \sim 10^{-10} \) , confirming a strong negative correlation between bump strength and redshift. This observation provides new insight into the nature and evolution of dust in quasar environments.
Although the ranges of bump strength and redshift are somewhat limited, and the completeness of the sample requires further verification, the observed trend still suggests a potential evolution in dust properties over cosmic time.

A large, homogeneous sample of quasar-associated 2175 \AA\ dust absorbers is essential for a robust statistical study of their properties. 
This will help address extinction biases related to quasar absorption lines and shed light on the physics of dust grains at different cosmic epochs and their relationships with other quasar absorbers.
Additionally, the study of BAL and NAL  systems  will enable measurements of the physical and chemical properties of the outflowing gas and the SM.
Direct exploration of multi-band magnitude data using machine learning algorithms, along with high-resolution spectroscopic follow-ups of known sources, will significantly enhance progress in both areas.
The exploration and study of quasar-associated 2175 \AA\ dust absorbers provides crucial insights into the interaction between dust, quasar activity, and galactic environments over cosmic time.

\begin{acknowledgments}
This work is supported by the National Natural Science Foundation of China (Grant No. 12173026 and 12141302), the National Key Research and Development Program of China (Grant No. 2022YFC2807303), and Shanghai Science and Technology Development Foundation (Grant No. 23010503900).
S.H.Z. acknowledges the support from the Program for Professor of Special Appointment (Eastern Scholar) at Shanghai Institutions of Higher Learning and the Shuguang Program (23SG39) of Shanghai Education Development Foundation and Shanghai Municipal Education Commission.  
Z.J.L. acknowledges the support from the Shanghai Science and Technology Foundation Fund (Grant No. 20070502400) and the scientific research grants from the China Manned Space Project. 
H.B.X. acknowledges the support from the National Natural Science Foundation of China (Grant No. 12203034) and the Shanghai Science and Technology Development Foundation (22YF1431500).
This work makes use of data from SDSS-IV. Funding for SDSS has been provided by the Alfred P. Sloan Foundation and Participating Institutions. Additional funding toward SDSS-IV has been provided by the U.S. Department of Energy Office of Science. SDSS-IV acknowledges support and resources from the Center for High-Performance Computing at the University of Utah. The SDSS website is www.sdss.org.
This publication makes use of data products from the Two Micron All Sky Survey, which is a joint project of the University of Massachusetts and the Infrared Processing and Analysis Center/California Institute of Technology, funded by the National Aeronautics and Space Administration and the National Science Foundation.
This publication makes use of data products from the Wide-field Infrared Survey Explorer, which is a joint project of the University of California, Los Angeles, and the Jet Propulsion Laboratory/California Institute of Technology, funded by the National Aeronautics and Space Administration.
This research is based on observations made with the Galaxy Evolution Explorer, obtained from the MAST data archive at the Space Telescope Science Institute, which is operated by the Association of Universities for Research in Astronomy, Inc., under NASA contract NAS 5-26555. 
\end{acknowledgments}

\end{CJK}
\end{document}